\RequirePackage{fix-cm}
\documentclass[smallextended]{svjour3}       
\smartqed  
\usepackage{graphicx}
\usepackage{xcolor}  
\usepackage{multirow}
\usepackage{amsmath}
\journalname{Quantum information processing}
\begin{document}

\title{Decoy-state round-robin differential-phase-shift quantum key distribution with source errors}
\thanks{The paper is supported by the National Natural Science Foundation of China (Grant No. 61871234, 61475075, 11847062) and the Postgraduate Research \& Practice Innovation Program of Jiangsu Province (Grant No. KYLX15-0832).}

\titlerunning{RRDPS-QKD with source errors}        

\author{Qian-Ping Mao \and Le Wang \and Sheng-Mei Zhao} 

\institute{Qian-Ping Mao
            \at Institute of Signal Processing and Transmission, Nanjing University of Posts and Telecommunications, Nanjing 210003, China
             \at College of computer science and technology, Nanjing Tech University, Nanjing 211816, China \\
              \email{maoqp@163.com}           
           \and
           Le Wang and Sheng-Mei Zhao  \at Institute of Signal Processing and Transmission, Nanjing University of Posts and Telecommunications, Nanjing 210003, China. \\
              \email{zhaosm@njupt.edu.cn}           
}

\date{Received: date / Accepted: date}

\maketitle

\begin{abstract}
As a promising quantum key distribution (QKD), most of the existing round-robin differential-phase-shift  quantum key distribution (RRDPS-QKD) protocols have adopted the decoy-state method and have assumed the source states are exactly controlled. However, the  precise manipulation of source states is impossible for any practical experiment, and the RRDPS-QKD with source errors has an unignorable impact on the performance of the protocol. In the paper, we study the four-intensity decoy-state RRDPS-QKD protocol with source errors, formulate the secure generation key rate of the proposed protocol, do the numerical simulations to testify the deductions. The results show that our evaluation can estimate the influence of source errors.

\keywords{Round-robin differential-phase-shift \and Quantum key distribution \and Decoy state \and Source errors}

\PACS{03.67.Dd \and 03.67.Hk}

\end{abstract}

\section{Introduction}
\label{intro}
Quantum key distribution (QKD) has developed rapidly due to its unconditional security \cite{Lo1999UnconditionalSecurityQuantum,Shor2000SimpleProofSecurity,Mayers2001Unconditionalsecurityquantum}. Since the fist QKD (BB84-QKD) protocol was proposed \cite{Bennett1984QuantumcryptographyPublic}, many more secure and practical QKD protocols have been developed, such as decoy-state QKD (DS-QKD) \cite{Lo2005DecoyStateQuantum,Wang2005BeatingPhotonNumber,Wang2009Decoystatequantum,Chi2012Decoystatemethod}, measurement-device-independent QKD (MDI-QKD) \cite{Lo2012MeasurementDeviceIndependent,Yin2016MeasurementDeviceIndependent}.

In order to guarantee the security, most of these protocols need to monitor the signal disturbance to estimate the amount of information leaked to Eve \cite{Gisin2002Quantumcryptography,Gottesman2004Securityquantumkey}. However, the round-robin differential-phase-shift QKD (RRDPS-QKD) \cite{Sasaki2014Practicalquantumkey}, put forward by Sasaki \emph{et al.} in 2014, can bound the information leakage without monitoring the signal disturbance, but depending on the signal states prepared by the sender. Moreover, with a large enough train length, the tolerance of bit error rate could be up to $50\%$ in theory.

Hence, as a promising and practical protocol, lots of efforts have been made to improve the performance and the practicality of the RRDPS-QKD \cite{Mizutani2015Robustnessroundrobin,Zhang2017Practicalroundrobin,Sasaki2017securityproofround,Yin2018Improvedsecuritybound,Zhang2016Practicalroundrobin,Wang2017Roundrobindifferential,Liu2017Roundrobindifferential,HuK2017RRDPS,Mao2017Plugplayround}.
Yin \emph{et al.} proposed an improved bound on information leakage to enhance the practicality and performance of RRDPS-QKD \cite{Yin2018Improvedsecuritybound}.
Zhang \emph{et al.} applied the tagging technique to overcome the effects of background noise and misalignment \cite{Zhang2017Practicalroundrobin}.
The heralded pair-coherent source (HPCS) \cite{Wang2017Roundrobindifferential} and the heralded single photon source (HSPS) \cite{Zhang2016Practicalroundrobin},
respectively, were adopted in the RRDPS-QKD protocol to improve the performance. We presented a plug-and-play RRDPS-QKD protocol to
make the RRDPS-QKD scheme be more practical \cite{Mao2017Plugplayround}.

Owing to the usage of the actual light source, the decoy-state method has been used in the most of existed RRDPS-QKD protocols to enhance the performance \cite{Zhang2017Practicalroundrobin,Zhang2016Practicalroundrobin,Wang2017Roundrobindifferential},
where the source states are always assumed to be perfectly controlled in the photon-number space.
However, it is impossible for any real experimental setup to keep the source states be
fully controlled, because of the fluctuations of source power, the environmental interference and other factors, etc.

On the other hand, the method to concern the source errors of decoy-states has been demonstrated in BB84-QKD and MDI-QKD protocols \cite{Wang2009Decoystatequantum,Wang2008Generaltheorydecoy,Wang2014Simulatingmeasurementdevice,Jiang2016Measurementdeviceindependent}.
Wang \emph{et al.} developed the general theory of decoy-state BB84-QKD with source errors
and studied the relationship between key generation rate and source errors \cite{Wang2009Decoystatequantum,Wang2008Generaltheorydecoy}.
With the intensity fluctuations, the formula of secret key generation rates for decoy-state BB84-QKD adopting the heralded single-photon source (HSPS) was calculated in ref.\cite{Wang2009Decoystatetheory},
and the heralded pair coherent source (HPCS) in ref.\cite{Zhou2010Decoystatequantum}.
The studies on the decoy-state MDI-QKD method with source errors were also presented in refs.\cite{Wang2014Simulatingmeasurementdevice,Jiang2016Measurementdeviceindependent}.
However, up to now, there has been no report on the decoy-state RRDPS-QKD method with source errors yet.

In this paper, we study the effect of source errors in the RRDPS-QKD protocol and show
the details of how to formulate the secure generation key rate by four-intensity decoy-state method.
We deduce the lower bound of count rate and the upper bound of bit error rate of
the k-photon state for the signal source only with one constraint of the source errors.
Moreover, we present the numerical simulations of according to our deductions.

\section{Four-intensity decoy-state RRDPS-QKD protocol with source errors}
\subsection{ The protocol}
\label{sec:2}
In the RRDPS-QKD protocol, Alice, the sender, prepares a train consisting of $L$ pulses, and encodes $L$ random bits on the phases of $L$ pulses in the train. Then the encoded pulse-train is sent to Bob through a quantum channel. Upon receiving the $L$-pulse train, Bob, the receiver, randomly picks two pulses in a train, and measures the relative phase between them by an Mach-Zehnder interferometer with a random delay. According to the publication of the successful interfered pulse-indices $(l, j)$, Alice can obtain her key bit according to the pulse-indices. These steps are repeated until Alice and Bob accumulate sufficient sifted key bits. After error correction and privacy amplification, Alice and Bob can finally share a secure key.

In practical implementations, weak coherent source (WCS) \cite{Zhang2017Practicalroundrobin}, HSPS \cite{Zhang2016Practicalroundrobin} or HPCS \cite{Wang2017Roundrobindifferential} are often adopted to replace the ideal single-photon source, thereafter, the decoy state method is employed to enhance the performance of the modified RRDPS-QKD protocol. Since the usage of infinite-intensity decoy-states is impractical, four-intensity decoy-state method is demonstrated to have the approaching performance \cite{Zhang2016Practicalroundrobin,Wang2017Roundrobindifferential} of the infinite-intensity decoy-state method.

In the four-intensity decoy-state RRDPS-QKD protocol, Alice has four source states, the signal state $\mu$ and three decoy states $\nu_1$, $\nu_2$ and $\nu_3$, and their density matrix can be written as $\rho = \sum\nolimits_{k=0}^J {p_{k,x} \left| k \right\rangle \left\langle k \right|}$ ($x=\mu, \nu_1, \nu_2, \nu_3$). Here, $J$ is determined by the type of light source, and it can be either finite or infinite. For WCS or HSPS, $J$ is $\infty$. The signal state is used to extract the final key, while the decoy states are used to estimate the lower bound of the gain and the upper bound of the bit error rate of the $k$-photon pulse-train. In the whole protocol, we assume that $M$ pulse-trains prepared by Alice is randomly chosen from one of the four states with the probability $P_x$ for $x=\mu, \nu_1, \nu_2, \nu_3$ and $P_\mu+P_{\nu_1}+P_{\nu_2}+P_{\nu_3}=1$.

Now, we consider the errors for the source states. At any $i$th ($i \in \{1,2,...,M\}$) pulse-train, each pulse in the train may have a slight deviation from the expected, and the state of the pulse-train
is expressed by  $\rho_i = \sum\nolimits_{k=0}^J {p_{ki,x} \left| k \right\rangle \left\langle k \right|}$,
where the $p_{ki,x}$ takes into account the errors of all the pulse in the train.
Without loss of generality, the minimum and maximum values of $p_{ki,x}$ are assumed to be bounded by $p_{k,x}^L$ and $p_{k,x}^U$, respectively, i.e., $p_{k,x}^L  \le p_{ki,x} \le p_{k,x}^U$, regardless of the error pattern.

In response to Alice, Bob observes the interference measurement for $M$ trains, and any successful interference result obtained from the $i$th pulse-train is called as "the $i$th train from Alice has caused a count" \cite{Wang2008Generaltheorydecoy}.
After completing all the measurements, Alice and Bob check the source state for each count by public discussion, and they obtain the number of count $N_x$ ($N_\mu$, $N_{\nu_1}$, $N_{\nu_2}$, $N_{\nu_3}$) and count rate caused from each source state $Q_x$($Q_\mu$, $Q_{\nu_1}$, $Q_{\nu_2}$, $Q_{\nu_3}$), where $Q_\mu=N_\mu/(P_\mu M)$,
$Q_{\nu_1}=N_{\nu_1}/(P_{\nu_1} M)$, $Q_{\nu_2}=N_{\nu_2}/(P_{\nu_2} M)$ and $Q_{\nu_3}=N_{\nu_3}/(P_{\nu_3} M)$.

\subsection{The secure key generation rate }
\label{sec:3}

The key generation rate per pulse of the RRDPS-QKD protocol can be given by \cite{Sasaki2014Practicalquantumkey}
\begin{equation}
R= \frac{1}{L}Q_{\mu}[1-fH_2(E_{b})-H_{PA}],
\end{equation}
where $Q_\mu$ is the overall gain when the average intensity of the pulse-trains is $\mu$, $E_{b}$ is the bit error rate, and $H_{PA}$ represents to the ratio of the key rate sacrificed in privacy amplification. $f$ denotes the efficiency of the error correction. $H_2(x)=-x\log_{2}(x)-(1-x)\log_{2}(1-x)$ is the information entropy function.

Here, $Q_\mu$, $Q_{\mu}E_{b}$, and $Q_{\mu}H_{PA}$ can be calculated by
$Q_\mu= \sum\nolimits_{k = 0}^\infty {Q_{k,\mu }}$,
$Q_\mu E_b= \sum\nolimits_{k = 0}^\infty {Q_{k,\mu } H_2(e_b^k)}$
and
$Q_\mu H_{PA}= \sum\nolimits_{k = 0}^\infty {Q_{k,\mu } H_2(e_{ph}^k)}$,
where $Q_{k,\mu}$, $e_{b}^k$ and $e_{ph}^k$ denote the count rate, the bit error rate and the phase rate for the k-photon state of the signal, respectively.
Note that, $Q_{k,\mu}$ and $e_{b}^k$ can not be measured directly, but they can be estimated by using decoy-states method.
Additionally, $e_{ph}^k$ is upper bounded by $e_{ph}^k \le k/(L-1)$ according to the RRDPS-QKD protocol \cite{Sasaki2014Practicalquantumkey,Zhang2017Practicalroundrobin}.

For practical implementations, the four-intensity decoy states is sufficient for the RRDPS-QKD protocol \cite{Zhang2016Practicalroundrobin,Wang2017Roundrobindifferential}.
The lower bounds of $Q_{0,\mu}$, $Q_{1,\mu}$, $Q_{2,\mu}$ ($Q_{0,\mu}^L$, $Q_{1,\mu}^L$, $Q_{2,\mu}^L$) can be estimated by using four-intensity decoy-states method, and the final key generation rate of our protocol can be lower bounded by
\begin{equation}
R \geq \frac{1}{L}\sum_{k=0}^{2}Q_{k,\mu}^L [1-H_2(e_{ph}^k)]-Q_{\mu} f H_2(E_{b}),
\end{equation}

Then, we show the estimations of $Q_{0,\mu}^L$, $Q_{1,\mu}^L$ and $Q_{2,\mu}^L$ in the following.

In order to estimate the low bound of $Q_{k,\mu}, k\in\{0,1,2\}$ , we also use the definition of set $C$ and set $c_k$  in ref.\cite{Wang2008Generaltheorydecoy}.

\emph{Definition}\cite{Wang2008Generaltheorydecoy}. The set $C$ contains any pulse-train that has caused a count, and the set $c_k$ contains any $k$-photon pulse-train that has caused a count. Mathematically, the necessary and sufficient condition for $i \in C$ is that the $i$th pulse-train has caused a count, and $i \in c_k$ is that the $i$th pulse-train containing $k$ photons has caused a count.

Accordingly, the overall counts of source state $x$ can be given
\begin{equation}
  N_x =\sum\limits_{k = 0}^J n_{k,x} = \sum\limits_{k = 0}^J \sum\limits_{i \in c_k } { \mathcal{P}_{xi|k}}
  =\sum\limits_{i \in c_0 } \mathcal{P}_{xi|0}  + \sum\limits_{i \in c_1 }  \mathcal{P}_{xi|1} + \sum\limits_{k = 2}^J {\sum\limits_{i \in c_k } { \mathcal{P}_{xi|k} } },
\end{equation}
where $n_{k,x}$ denotes the number of counts caused by those pulse-trains containing $k$ photons from  source state $x$ ($x=\mu, \nu_1, \nu_2, \nu_3$), and $\mathcal{P}_{xi|k}$ is the probability of the $i$th pulse-train from the source state $x$ on the premise of the $i$th pulse-train containing $k$ photons. Here, $J$ can be either finite or infinite.

According to the protocol, $\mathcal{P}_{xi|k}$ is given by
\begin{equation}
  \mathcal{P}_{xi|k} = \frac{{P_x \cdot p_{ki,x} }}{{P_\mu \cdot p_{ki,\mu } + P_{\nu _1 }\cdot p_{ki,\nu _1 } + P_{\nu _2 }  \cdot p_{ki,\nu _2} + P_{\nu_3 } \cdot p_{ki,\nu_3 } }}.
\end{equation}
Let $d_{ki} = 1/(P_\mu \cdot p_{ki,\mu }  + P_{\nu _1 } \cdot p_{ki,\nu _1 }  + P_{\nu _2 } \cdot p_{ki,\nu _2} + P_{\nu _3 }  \cdot p_{ki,\nu _3 } )$, we can have
\begin{equation}
  \mathcal{P}_{xi|k}=P_x \cdot p_{ki,x} \cdot d_{ki}
\end{equation}
and
\begin{equation}
  N_x = \sum\limits_{k = 0}^J n_{k,x}= \sum\limits_{k = 0}^J \sum\limits_{i \in c_k } P_x \cdot p_{ki,x} \cdot d_{ki},
\end{equation}
i.e.,
\begin{eqnarray} 
  \frac{N_\mu}{P_\mu}&=& \sum\limits_{i \in c_0 } p_{0i,\mu} d_{0i}+ \sum\limits_{i \in c_1 } p_{1i,\mu} d_{1i}+  \sum\limits_{k = 2}^J \sum\limits_{i \in c_k } p_{ki,\mu} d_{ki}, \quad \quad \label{Nu}\\
  \frac{N_{\nu_1}}{P_{\nu_1}}&=& \sum\limits_{i \in c_0 } p_{0i,\nu_1} d_{0i}+ \sum\limits_{i \in c_1 } p_{1i,\nu_1} d_{1i}+ \sum\limits_{k = 2}^J \sum\limits_{i \in c_k } p_{ki,\nu_1} d_{ki}, \label{Nv1} \\
  \frac{N_{\nu_2}}{P_{\nu_2}}&=& \sum\limits_{i \in c_0 } p_{0i,\nu_2} d_{0i}+ \sum\limits_{i \in c_1 } p_{1i,\nu_2} d_{1i}+ \sum\limits_{k = 2}^J \sum\limits_{i \in c_k } p_{ki,\nu_2} d_{ki},  \label{Nv2}\\
  \frac{N_{\nu_3}}{P_{\nu_3}}&=& \sum\limits_{i \in c_0 } p_{0i,\nu_3} d_{0i}+ \sum\limits_{i \in c_1 } p_{1i,\nu_3} d_{1i}+ \sum\limits_{k = 2}^J \sum\limits_{i \in c_k } p_{ki,\nu_3} d_{ki}. \label{Nv3}
\end{eqnarray} 

If the source errors exist, $n_{k,\mu}$, $n_{k,\nu_1}$, $n_{k,\nu_2}$ and $n_{k,\nu_3}$ should be bounded by
\begin{eqnarray}
  p_{k,\mu }^L \sum\limits_{i \in c_k } {d_{ki} }  \le  \frac{n_{k,\mu}}{ P_\mu} &=&  \sum\limits_{i \in c_k } {p_{ki,\mu } d_{ki} }  \le p_{k,\mu }^U \sum\limits_{i \in c_k } {d_{ki} },\quad \quad\\
  p_{k,\nu _1 }^L \sum\limits_{i \in c_k } {d_{ki} } \le \frac{n_{k,\nu _1}}{P_{\nu _1 }} &=& \sum\limits_{i \in c_k } {p_{ki,\nu _1 } d_{ki} }  \le p_{k,\nu _1 }^U \sum\limits_{i \in c_k } {d_{ki} },\\
  p_{k,\nu _2 }^L \sum\limits_{i \in c_k } {d_{ki} } \le \frac{n_{k,\nu _2}}{P_{\nu _2 }}  &=& \sum\limits_{i \in c_k } {p_{ki,\nu _2} d_{ki} }  \le p_{k,\nu _2 }^U \sum\limits_{i \in c_k } {d_{ki} },\\
  p_{k,\nu _3 }^L \sum\limits_{i \in c_k } {d_{ki} } \le \frac{n_{k,\nu _3}}{P_{\nu _3}} &=& \sum\limits_{i \in c_k } {p_{ki,\nu _3 } d_{ki} }  \le p_{k,\nu _3 }^U \sum\limits_{i \in c_k } {d_{ki} }.
\end{eqnarray}
Further, the important conditions
\begin{eqnarray}
 & & \frac{{p_{k,\mu}^L }}{{p_{k,\nu _1 }^U }} \ge \frac{{p_{2,\mu }^L }}{{p_{2,\nu _1 }^U }} \ge \frac{{p_{1,\mu}^L }}{{p_{1,\nu _1 }^U }}  \ge \frac{{p_{0,\mu}^L }}{{p_{0,\nu _1 }^U }}, \quad for\;all\;k \ge 2 \nonumber \\
 & & \frac{{p_{k,\nu _1 }^L }}{{p_{k,\nu _2 }^U }} \ge \frac{{p_{2,\nu _1 }^L }}{{p_{2,\nu _2 }^U }} \ge \frac{{p_{1,\nu _1 }^L }}{{p_{1,\nu _2 }^U }}  \ge \frac{{p_{0,\nu _1 }^L }}{{p_{0,\nu _2 }^U }}, \quad for\;all\;k \ge 2 \nonumber \\
 & & \frac{{p_{k,\nu _2 }^L }}{{p_{k,\nu _3 }^U }} \ge \frac{{p_{2,\nu _2 }^L }}{{p_{2,\nu _3 }^U }} \ge \frac{{p_{1,\nu _2 }^L }}{{p_{1,\nu _3 }^U }} \ge \frac{{p_{0,\nu _2 }^L }}{{p_{0,\nu _3 }^U }}, \quad for\;all\;k \ge 2 \label{Condition}
\end{eqnarray}
need to be satisfied, which has been demonstrated in refs.\cite{Wang2008Generaltheorydecoy,Wang2009Decoystatetheory,Zhou2010Decoystatequantum}.

To minimize the values of $Q_{0,\mu}$, $Q_{1,\mu}$ and $Q_{2,\mu}$, we need to minimize $D_0$, $D_1$ and $D_2$, which are defined as $D_0=\sum\nolimits_{i \in c_k }d_{0i}$, $D_1=\sum\nolimits_{i \in c_k }d_{1i}$, $D_2=\sum\nolimits_{i \in c_k }d_{2i}$.
Using Eqs.(\ref{Nu})-(\ref{Nv3}), the lower bound of $D_0$, $D_1$, $D_2$ can be estimated as the following, (see the details in the Appendix)
\begin{eqnarray}
  D_0^L &=& \max \left\{ {\frac{{p_{1,\nu _1 }^L \frac{{N_{\nu _2 } }}{{P_{\nu _2 } }} - p_{1,\nu _2 }^U \frac{{N_{\nu _1 } }}{{P_{\nu _1 } }}}}{{p_{1,\nu _1 }^L p_{0,\nu _2 }^U  - p_{1,\nu _2 }^U p_{0,\nu _1 }^L }},\,0} \right\}, \nonumber
  \\
  D_1^L &=& \frac{{(p_{0,\nu _2 }^L \frac{{N_{\nu _1 } }}{{P_{\nu _1 } }} - p_{0,\nu _1 }^U \frac{{N_{\nu _2 } }}{{P_{\nu _2 } }})p_{2,\mu }^L  - q_1(\frac{{N_\mu  }}{{P_\mu  }} - p_{0,\mu }^L D_0^L )}}{{q_3 p_{2,\mu }^L  - q_1 p_{1,\mu }^L }}, \nonumber
  \\
  D_2^L &=& \bigg\{ \bigg[(p_{0,\nu _2 }^L \frac{{N_{\nu _1 } }}{{P_{\nu _1 } }} - p_{0,\nu _1 }^U \frac{{N_{\nu _2 } }}{{P_{\nu _2 } }}) q_2 -(p_{0,\nu _3 }^U \frac{{N_{\nu _2 } }}{{P_{\nu _2 } }} - p_{0,\nu _2 }^L \frac{{N_{\nu _3 } }}{{P_{\nu _3 } }}) q_3\bigg]p_{3,\mu }^L
  \nonumber \\
  & & -(s_1 q_2 - s_2 q_3)(\frac{{N_\mu  }}{{P_\mu  }} - p_{0,\mu }^L D_0^L  - p_{1,\mu }^L D_1^L ) \bigg\}
  \nonumber \\
  & & \bigg/ \bigg\{ \bigg[q_1 q_2 - (p_{0,\nu _3 }^U p_{2,\nu _2 }^L  - p_{0,\nu _2 }^L p_{2,\nu _3 }^U )q_3 \bigg] p_{3,\mu }^L - (s_1 q_2 - s_2 q_3) p_{2,\mu }^L \bigg\},
\end{eqnarray}
where $q_1 = p_{0,\nu _2 }^L p_{2,\nu _1 }^U  - p_{0,\nu _1 }^U p_{2,\nu _2 }^L $,
$q_2 = p_{0,\nu _3 }^U p_{1,\nu _2 }^L  - p_{0,\nu _2 }^L p_{1,\nu _3 }^U$,
$q_3 = p_{0,\nu _2 }^L p_{1,\nu _1 }^U  - p_{0,\nu _1 }^U p_{1,\nu _2 }^L $,
$s_1 = p_{0,\nu _2 }^L p_{3,\nu _1 }^U  - p_{0,\nu _1 }^U p_{3,\nu _2 }^L$,
$s_2 = p_{0,\nu _3 }^U p_{3,\nu _2 }^L  - p_{0,\nu _2 }^L p_{3,\nu _3 }^U$.
Therefore, according to the definition of $Q_{k,\mu }^L  = {n_{k,\mu }}/(P_\mu  M) = {p_{k,\mu }^L D_k^L }/{M}$, the lower bounds of $Q_{0,\mu}$, $Q_{1,\mu}$ and $Q_{2,\mu}$ are given by
\begin{small}
\begin{eqnarray}
  Q_{0,\mu}^L &\!=\!& \max \left\{ {\frac{{p_{0,\mu }^L (p_{1,\nu _1 }^L Q_{\nu_2 }- p_{1,\nu_2 }^U Q_{\nu_1})}}{{p_{1,\nu _1 }^L p_{0,\nu_2 }^U - p_{1,\nu_2 }^U p_{0,\nu_1 }^L }},\,0} \right\}, \\
 Q_{1,\mu}^L &\!=\!& \frac{{p_{1,\mu}^L \left[ {(p_{0,\nu _2 }^L Q_{\nu _1 }  - p_{0,\nu _1 }^U Q_{\nu _2 } )p_{2,\mu }^L  - q_1 (Q_\mu- Q_{0,\mu }^L )} \right]}}{{q_3 p_{2,\mu }^L  - q_1 p_{1,\mu }^L }}, \nonumber
 \\
 Q_{2,\mu}^L &\!=\!& \frac{p_{2,\mu}^L\bigg\{\bigg[(p_{0,\nu_2}^L Q_{\nu_1} - p_{0,\nu_1}^U Q_{\nu_2}) q_2 - (p_{0,\nu_3}^U Q_{\nu_2 }-p_{0,\nu_2}^L Q_{\nu_3})q_3\bigg]p_{3,\mu }^L-(s_1 q_2 - s_2 q_3)(Q_\mu  - Q_{0,\mu}^L-Q_{1,\mu}^L ) \bigg\}}{\bigg\{ \bigg[q_1 q_2 - (p_{0,\nu_3}^U p_{2,\nu_2}^L - p_{0,\nu_2}^L p_{2,\nu_3}^U )q_3\bigg]p_{3,\mu}^L-(s_1 q_2-s_2 q_3) p_{2,\mu}^L \bigg\}}.\nonumber
\end{eqnarray}
\end{small}

\section{Numerical simulations}
\label{sec:6}
Utilizing all the above deduction, we can conclude the secure key generation rate of the proposed QKD protocol. Here, we use WCS source as an example, we calculate and present the secure key rate performance of the proposed QKD protocol.

For WCS source with intensity $x$ per pulse-train, it follows a Poisson distribution, and the k-photon probability could be given by $p_{k,x}= e^{-x}x^{k}/{k!}$.
Considering the source errors, we assume that the maximum deviation of the intensity per pulse of the source x is $\delta_{x}$ ($x=\mu, \nu_1, \nu_2$, $\nu_3$). Hence the intensity of the train $i$ could be bounded by $x(1 - \delta_x ) \le x_i \le x(1+\delta_x )$. Note that the ${\nu_1}_i \geq{\nu_2}_i \geq{\nu_3}_i \geq0$ and $({\nu_1}_i +{\nu_2}_i +{\nu_3}_i )<\mu_i<1$ need to be guaranteed in our protocol according to the conditions of Eq.(\ref{Condition}).
For simplicity, we assume all the signal source state and the three decoy source states have the same upper bound $\delta$, i.e., $\delta=\delta_\mu=\delta_{\nu_1}=\delta_{\nu_2}=\delta_{\nu_3}$.
Hence, $p_{ki,x}$ is bounded by $[p_{k,x}^L, p_{k,x}^U]$, and
\begin{eqnarray}
p_{0,x}^L=p_{0,x(1+\delta )}, \quad   p_{0,x}^U=p_{0,x(1-\delta )}, \nonumber \\
p_{k,x}^L=p_{k,x(1-\delta )}, \quad  p_{k,x}^U=p_{k,x(1+\delta )},
\end{eqnarray}
for all $k \ge 1$.

In addition, the overall count rate $Q_x$ ($Q_\mu$, $Q_{\nu_1}$, $Q_{\nu_2}$, $Q_{\nu_3}$) and the overall QBER ($E_\mu$, $E_{\nu_1}$, $E_{\nu_2}$, $E_{\nu_3}$), which can be directly measured in practical experiments, here, are given by
\begin{eqnarray}
  Q_{\mu}&=& 1-(1-p_{d})e^{-\mu \eta _{t} \eta _{B}}, \nonumber \\
  Q_{\nu_1}&=& 1-(1-p_{d})e^{-\nu_1 \eta _{t} \eta _{B}},\nonumber \\
  Q_{\nu_2}&=& 1-(1-p_{d})e^{-\nu_2 \eta _{t} \eta _{B}},\nonumber \\
  Q_{\nu_3}&=& 1-(1-p_{d})e^{-\nu_3 \eta _{t} \eta _{B}},
\end{eqnarray}
and
\begin{eqnarray}
  E_{\mu}&=& [e_{d}(1-p_{d})(1-e^{-\mu \eta _{t} \eta _{B}})+e_{0}p_{d}]/Q_{\mu},\nonumber \\
  E_{\nu_1}&=& [e_{d}(1-p_{d})(1-e^{-\nu_1 \eta _{t} \eta _{B}})+e_{0}p_{d}]/Q_{\nu_1},\nonumber \\
  E_{\nu_2}&=& [e_{d}(1-p_{d})(1-e^{-\nu_2 \eta _{t} \eta _{B}})+e_{0}p_{d}]/Q_{\nu_2},\nonumber \\
  E_{\nu_3}&=& [e_{d}(1-p_{d})(1-e^{-\nu_3 \eta _{t} \eta _{B}})+e_{0}p_{d}]/Q_{\nu_3},
\end{eqnarray}
where $e_{0}$ and $e_{d}$ are the error probabilities caused by the background and the misalignment, respectively. $p_{d}$ denotes the background count rate for the detector, and $\eta _{B}$ is the efficiency of Bob's detectors. $\eta _{t}$ is the efficiency of the channel transmission, which is expressed as $\eta _{t}=10^{-\alpha z/10}$,
where $\alpha$ and $z$ are the channel transmission loss rate and the transmission distance, respectively.

In our numerical simulations, the corresponding parameters are listed in Tab.\ref{tab:1} \cite{Gobby2004Quantumkeydistribution},
where $L$ represents the pulse train length. Furthermore, the values of $\mu$, $\nu_1$, $\nu_2$ and $\nu_3$ are optimized to obtain the optimal key rate performance.

\begin{table}[htbp]
\centering
\caption{Parameters for our simulations}
\label{tab:1}
\begin{tabular}{cccccc}
\hline\noalign{\smallskip}
$P_{d}$ & $e_{d}$ & $e_0$ & $\eta_{B}$ & $\alpha$ & $f$ \\
\noalign{\smallskip}\hline\noalign{\smallskip}
$1.7\times10^{-6}L$ & $3.3\%$ & $50\%$ & $4.5\%$ & $0.2dB/km$ & $1.16$ \\
\noalign{\smallskip}\hline
\end{tabular}
\end{table}

\begin{figure}[htbp]
\centering
\includegraphics [width=0.7\columnwidth]{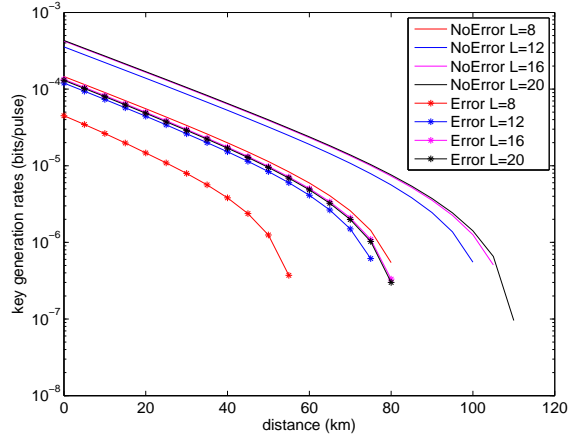} 
\caption{Key rate comparison for the RRDPS-QKD with and without source errors.}
\label{fig:1}
\end{figure}

Fig.\ref{fig:1} shows the performance of the RRDPS-QKD protocol with source errors and without source errors for $L=8$, $L=12$, $L=16$ and $L=20$. Here, parameter $\delta$ of the source errors is set as $0.05$.
The result shows that, for the protocol with source errors, the key rates decrease with the increasing transmission distance,
and the performance for $L=16$ and $L=20$ is better. Moreover, from the comparison between the protocol with and without source errors, it is found that the performance of the protocol with source errors is always worse than the case without source errors.

\begin{figure}[htbp]
\centering
\includegraphics [width=0.7\columnwidth]{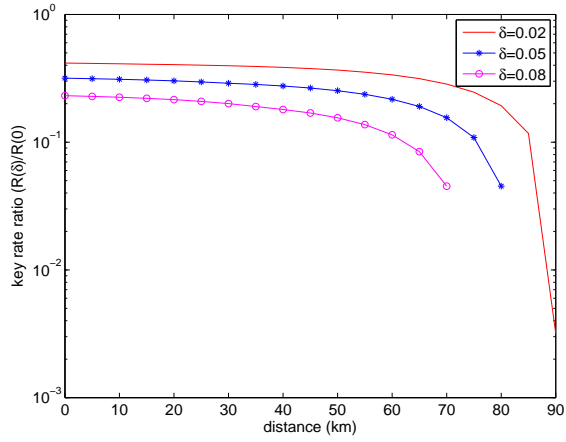} 
\caption{Key rate ratio versus transmission distance for different $\delta$.}
\label{fig:2}
\end{figure}

\begin{figure}[htbp]
\centering
\includegraphics [width=0.7\columnwidth]{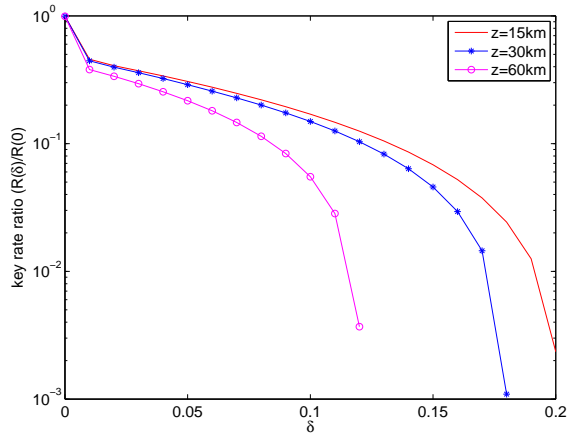} 
\caption{Key rate ratio against $\delta$ for different transmission distance $z$.}
\label{fig:3}
\end{figure}

The influence of source errors on the key rate is presented in Fig.\ref{fig:2} and Fig.\ref{fig:3}, where $L$ is set to $16$.

Fig.\ref{fig:2} shows the key rate ratios ($R(\delta)/R(0)$) versus transmission distance for $\delta=0.02$, $0.05$ and $0.08$, where $R(\delta)$ is the key rate with source errors and $R(0)$ is the key rate without source errors. From Fig.\ref{fig:2}, it can be seen that $\delta$ has an effect on the key rate, especially over a large distance. Moreover, the larger $\delta$ is, the smaller the value of $R(\delta)/R(0)$ is, i.e., the greater the impact on key rate is.

Fig.\ref{fig:3} depicts the key rate ratios ($R(\delta)/R(0)$) against $\delta$ for transmission distance $z=15$km, $30$km and $60$km. The results indicate that the influence on the key rates increases with the increasing $\delta$ for any certain transmission distance. Comparing with the key rate ratios over three transmission distances, it is found that the larger transmission distance is, the smaller the ratio on key rate is, which means the greater influence.

\section{Conclusions}
In this paper, we have analyzed the influence of source errors on the secure key generation rate of the RRDPS-QKD protocol with four-intensity decoy-states. We have presented tight bounds of the key generation rate when the count rate and the bit error rate of k-photon states for the signal state are bounded.  WCS has been used as the example to discuss the performance of the four-intensity decoy-state RRDPS-QKD protocol with source errors. The results have shown that the practical RRDPS-QKD with source errors has an unignorable impact on the secure key generation rate. The larger source error is, the higher degradation of secure key rate is. Simultaneously, the longer transmission distance is, the greater secure key rate degrades.
By extending our formulas to any type of source, such as HSPS and HPCS, the influence on the performance of practical RRDPS-QKD protocol with source errors can be estimated.

\begin{acknowledgements}
The paper is supported by the National Natural Science Foundation of China (Grant No. 61871234, 61475075, 11847062) and the Postgraduate Research \& Practice Innovation Program of Jiangsu Province (Grant No. KYLX15-0832).
\end{acknowledgements}

\section{Appendix}
In this section, we will give the lower bounds of $D_0$,$D_1$,$D_2$.

In order to get the lower bound of $D_0$, applying $p_{1,\nu _1 }^L Eq.(\ref{Nv2}) -p_{1,\nu _2 }^U Eq.(\ref{Nv1})$, we have
\begin{small}
\begin{eqnarray}
 & & p_{1,\nu _1 }^L\frac{{N_{\nu _2 } }}{{P_{\nu _2 } }} - p_{1,\nu _2 }^U \frac{{N_{\nu _1 } }}{{P_{\nu _1 } }}
 \nonumber \\
= & &\sum\limits_{i \in c_0 } {\left( {p_{1,\nu _1 }^L p_{0i,\nu _2 }  - p_{1,\nu _2 }^U p_{0i,\nu _1 } } \right)d_{0i} } + \sum\limits_{i \in c_1 } {\left( {p_{1,\nu _1 }^L p_{1i,\nu _2 }  - p_{1,\nu _2 }^U p_{1i,\nu _1 } } \right)d_{1i} }
 \nonumber \\
 & & \quad \quad \quad \quad + \sum\limits_{k = 2}^J {\sum\limits_{i \in c_k } {\left( {p_{1,\nu _1 }^L p_{ki,\nu _2 }  - p_{1,\nu _2 }^U p_{ki,\nu _1 } } \right)d_{ki} } }
 \nonumber \\
 \le & & \sum\limits_{i \in c_0 } {\left( {p_{1,\nu _1 }^L p_{0,\nu _2 }^U  - p_{1,\nu _2 }^U p_{0,\nu _1 }^L } \right)d_{0i} } + \sum\limits_{i \in c_1 } {\left( {p_{1,\nu _1 }^L p_{1,\nu _2 }^U  - p_{1,\nu _2 }^U p_{1,\nu _1 }^L } \right)d_{1i} }
  \nonumber \\
 & & \quad \quad \quad \quad  + \sum\limits_{k = 2}^J {\sum\limits_{i \in c_k } {\left( {p_{1,\nu _1 }^L p_{k,\nu _2 }^U  - p_{1,\nu _2 }^U p_{k,\nu _1 }^L } \right)d_{ki} } }
 \nonumber \\
=& &\left( {p_{1,\nu _1 }^L p_{0,\nu _2 }^U  - p_{1,\nu _2 }^U p_{0,\nu _1 }^L } \right)\sum\limits_{i \in c_0 } {d_{0i} }+ \sum\limits_{k = 2}^J \left( {p_{1,\nu _1 }^L p_{k,\nu _2 }^U  - p_{1,\nu _2 }^U p_{k,\nu _1 }^L } \right) {\sum\limits_{i \in c_k } {d_{ki} } }. \label{InqD00}
\end{eqnarray}
\end{small}
According to the definitions of $D_0$, $D_1$ and $D_2$, we get
\begin{eqnarray}
 & & p_{1,\nu _1 }^L\frac{{N_{\nu _2 } }}{{P_{\nu _2 } }} - p_{1,\nu _2 }^U \frac{{N_{\nu _1 } }}{{P_{\nu _1 } }}
 \nonumber \\
 \le & & \left( {p_{1,\nu _1 }^L p_{0,\nu _2 }^U  - p_{1,\nu _2 }^U p_{0,\nu _1 }^L } \right) D_0+\sum\limits_{k = 2}^J \left( {p_{1,\nu _1 }^L p_{k,\nu _2 }^U  - p_{1,\nu _2 }^U p_{k,\nu _1 }^L } \right) {D_k}. \label{InqD01}
\end{eqnarray}
Due to the conditions of Eq.(\ref{Condition}),
we can obtain an inequality $p_{1,\nu _1 }^L p_{k,\nu _2 }^U  - p_{1,\nu _2 }^U p_{k,\nu _1 }^L  \le 0 \quad for\;all\;k \ge 2$.
Therefore, the inequality in Eq.(\ref{InqD01}) becomes
\begin{equation}
 p_{1,\nu _1 }^L\frac{{N_{\nu _2 } }}{{P_{\nu _2 } }} - p_{1,\nu _2 }^U \frac{{N_{\nu _1 } }}{{P_{\nu _1 } }}
 \le  \left( {p_{1,\nu _1 }^L p_{0,\nu _2 }^U  - p_{1,\nu _2 }^U p_{0,\nu _1 }^L } \right) D_0 ,
\end{equation}
and the lower bound of $D_0$ is obtained by
\begin{equation}
 D_0  \ge D_0^L  = \max \left\{ {\frac{{p_{1,\nu _1 }^L \frac{{N_{\nu _2 } }}{{P_{\nu _2 } }} - p_{1,\nu _2 }^U \frac{{N_{\nu _1 } }}{{P_{\nu _1 } }}}}{{p_{1,\nu _1 }^L p_{0,\nu _2 }^U  - p_{1,\nu _2 }^U p_{0,\nu _1 }^L }},\,0} \right\}. \label{InqD02}
\end{equation}

Then, similar to Eq.(\ref{InqD00}), by $p_{0,\nu _2 }^L Eq.(\ref{Nv1}) -p_{0,\nu _1 }^U Eq.(\ref{Nv2})$, we have
\begin{small}
\begin{eqnarray}
 & & p_{0,\nu _2 }^L \frac{{N_{\nu _1 } }}{{P_{\nu _1 } }} - p_{0,\nu _1 }^U \frac{{N_{\nu _2 } }}{{P_{\nu _2 } }}
 \nonumber \\
 \le & &(p_{0,\nu _2 }^L p_{1,\nu _1 }^U  - p_{0,\nu _1 }^U p_{1,\nu _2 }^L )D_1 + \sum\limits_{k = 2}^J {\sum\limits_{i \in c_k } {(p_{0,\nu _2 }^L p_{k,\nu _1 }^U  - p_{0,\nu _1 }^U p_{k,\nu _2 }^L )d_{ki} } }
 \nonumber \\
  \le & & (p_{0,\nu _2 }^L p_{1,\nu _1 }^U  - p_{0,\nu _1 }^U p_{1,\nu _2 }^L )D_1+ \sum\limits_{k = 2}^J \frac{p_{0,\nu _2 }^L p_{k,\nu _1 }^U  - p_{0,\nu _1 }^U p_{k,\nu _2 }^L }{{p_{k,\mu }^L }} {\sum\limits_{i \in c_k } {d_{ki} p_{ki,\mu } } }
  \nonumber \\
  \le & & (p_{0,\nu _2 }^L p_{1,\nu _1 }^U - p_{0,\nu _1 }^U p_{1,\nu _2 }^L )D_1 + \frac{p_{0,\nu _2 }^L p_{2,\nu _1 }^U  - p_{0,\nu _1 }^U p_{2,\nu _2 }^L }{{p_{2,\mu }^L }}[\frac{{N_\mu  }}{{P_\mu  }} - p_{0,\mu }^L D_0^L - p_{1,\mu }^L D_1 ]. \label{InqD11}
\end{eqnarray}
\end{small}
Here, the inequality that $\frac{(p_{0,\nu _2 }^L p_{k,\nu _1 }^U  - p_{0,\nu _1 }^U p_{k,\nu _2 }^L )}{{p_{k,\mu }^L }} \le \frac{(p_{0,\nu _2 }^L p_{2,\nu _1 }^U  - p_{0,\nu _1 }^U p_{2,\nu _2 }^L )}{{p_{2,\mu }^L }}$ for $k \ge 2$ is adopted to prove the inequality in Eq.(\ref{InqD11}).
Consequently, the lower bound of $D_1$ can be estimated by
\begin{small}
\begin{equation}
D_1  \ge D_1^L  = \frac{{(p_{0,\nu _2 }^L \frac{{N_{\nu _1 } }}{{P_{\nu _1 } }} - p_{0,\nu _1 }^U \frac{{N_{\nu _2 } }}{{P_{\nu _2 } }})p_{2,\mu }^L  - (p_{0,\nu _2 }^L p_{2,\nu _1 }^U  - p_{0,\nu _1 }^U p_{2,\nu _2 }^L )(\frac{{N_\mu  }}{{P_\mu  }} - p_{0,\mu }^L D_0^L )}}{{(p_{0,\nu _2 }^L p_{1,\nu _1 }^U  - p_{0,\nu _1 }^U p_{1,\nu _2 }^L )p_{2,\mu }^L  - (p_{0,\nu _2 }^L p_{2,\nu _1 }^U  - p_{0,\nu _1 }^U p_{2,\nu _2 }^L )p_{1,\mu }^L }},  \label{InqD12}
\end{equation}
\end{small}
where $D_0^L$ has been calculated in Eq.(\ref{InqD02}).

Combining the two equations ($p_{0,\nu_2}^L Eq.(\ref{Nv1})-p_{0,\nu_1}^U Eq.(\ref{Nv2}) $) and ($p_{0,\nu_3}^U Eq.(\ref{Nv2})-p_{0,\nu_2 }^L Eq.(\ref{Nv3})$), we can have
\begin{small}
\begin{eqnarray}
 & & (p_{0,\nu _2 }^L \frac{{N_{\nu _1 } }}{{P_{\nu _1 } }} - p_{0,\nu _1 }^U \frac{{N_{\nu _2 } }}{{P_{\nu _2 } }})(p_{0,\nu _3 }^U p_{1,\nu _2 }^L  - p_{0,\nu _2 }^L p_{1,\nu _3 }^U ) - (p_{0,\nu _3 }^U \frac{{N_{\nu _2 } }}{{P_{\nu _2 } }} - p_{0,\nu _2 }^L \frac{{N_{\nu _3 } }}{{P_{\nu _3 } }})(p_{0,\nu _2 }^L p_{1,\nu _1 }^U  - p_{0,\nu _1 }^U p_{1,\nu _2 }^L )
 \nonumber \\
 &\le &  [(p_{0,\nu _2 }^L p_{2,\nu _1 }^U  - p_{0,\nu _1 }^U p_{2,\nu _2 }^L )(p_{0,\nu _3 }^U p_{1,\nu _2 }^L  - p_{0,\nu _2 }^L p_{1,\nu _3 }^U ) - (p_{0,\nu _3 }^U p_{2,\nu _2 }^L  - p_{0,\nu _2 }^L p_{2,\nu _3 }^U )(p_{0,\nu _2 }^L p_{1,\nu _1 }^U  - p_{0,\nu _1 }^U p_{1,\nu _2 }^L )]D_2
 \nonumber \\
 & & +  \sum\limits_{k = 3}^J \frac{{(p_{0,\nu _2 }^L p_{k,\nu _1 }^U  - p_{0,\nu _1 }^U p_{k,\nu _2 }^L )(p_{0,\nu _3 }^U p_{1,\nu _2 }^L  - p_{0,\nu _2 }^L p_{1,\nu _3 }^U ) - (p_{0,\nu _3 }^U p_{k,\nu _2 }^L  - p_{0,\nu _2 }^L p_{k,\nu _3 }^U )(p_{0,\nu _2 }^L p_{1,\nu _1 }^U  - p_{0,\nu _1 }^U p_{1,\nu _2 }^L )}}{{p_{k,\mu }^L }}{\sum\limits_{i \in c_k } {p_{ki,\mu } d_{ki} } }
  \nonumber \\
  & \le & [(p_{0,\nu _2 }^L p_{2,\nu _1 }^U  - p_{0,\nu _1 }^U p_{2,\nu _2 }^L )(p_{0,\nu _3 }^U p_{1,\nu _2 }^L  - p_{0,\nu _2 }^L p_{1,\nu _3 }^U ) - (p_{0,\nu _3 }^U p_{2,\nu _2 }^L  - p_{0,\nu _2 }^L p_{2,\nu _3 }^U )(p_{0,\nu _2 }^L p_{1,\nu _1 }^U  - p_{0,\nu _1 }^U p_{1,\nu _2 }^L )]D_2
  \nonumber  \\
  & &+ \frac{{(p_{0,\nu _2 }^L p_{3,\nu _1 }^U  - p_{0,\nu _1 }^U p_{3,\nu _2 }^L )(p_{0,\nu _3 }^U p_{1,\nu _2 }^L  - p_{0,\nu _2 }^L p_{1,\nu _3 }^U ) - (p_{0,\nu _3 }^U p_{3,\nu _2 }^L - p_{0,\nu _2 }^L p_{3,\nu _3 }^U )(p_{0,\nu _2 }^L p_{1,\nu _1 }^U  - p_{0,\nu _1 }^U p_{1,\nu _2 }^L )}}{{p_{3,\mu }^L }}  \nonumber  \\
 & & \quad \quad  \quad \quad \quad \quad \cdot (\frac{{N_\mu}}{{P_\mu}} - p_{0,\mu }^L D_0^L - p_{1,\mu }^L D_1^L  - p_{2,\mu }^L D_2 ),
\end{eqnarray}
\end{small}
where the inequality is due to $\frac{{(p_{0,\nu _2 }^L p_{k,\nu _1 }^U  - p_{0,\nu _1 }^U p_{k,\nu _2 }^L )(p_{0,\nu _3 }^U p_{1,\nu _2 }^L  - p_{0,\nu _2 }^L p_{1,\nu _3 }^U ) - (p_{0,\nu _3 }^U p_{k,\nu _2 }^L  - p_{0,\nu _2 }^L p_{k,\nu _3 }^U )(p_{0,\nu _2 }^L p_{1,\nu _1 }^U  - p_{0,\nu _1 }^U p_{1,\nu _2 }^L )}} {{p_{k,\mu }^L }} \le \frac{{(p_{0,\nu _2 }^L p_{3,\nu _1 }^U  - p_{0,\nu _1 }^U p_{3,\nu _2 }^L )(p_{0,\nu _3 }^U p_{1,\nu _2 }^L  - p_{0,\nu _2 }^L p_{1,\nu _3 }^U ) - (p_{0,\nu _3 }^U p_{3,\nu _2 }^L - p_{0,\nu _2 }^L p_{3,\nu _3 }^U )(p_{0,\nu _2 }^L p_{1,\nu _1 }^U  - p_{0,\nu _1 }^U p_{1,\nu _2 }^L )}} {{p_{3,\mu }^L }} $ for $k\ge3$.

Hence, the lower bound of $D_2$ can be expressed by
\begin{small}
\begin{eqnarray}
& & D_2\ge D_2^L
 \nonumber \\
&=& \bigg\{\bigg[(p_{0,\nu _2 }^L \frac{{N_{\nu _1 } }}{{P_{\nu _1 } }} - p_{0,\nu _1 }^U \frac{{N_{\nu _2 } }}{{P_{\nu _2 } }})(p_{0,\nu _3 }^U p_{1,\nu _2 }^L  - p_{0,\nu _2 }^L p_{1,\nu _3 }^U ) - (p_{0,\nu _3 }^U \frac{{N_{\nu _2 } }}{{P_{\nu _2 } }} - p_{0,\nu _2 }^L \frac{{N_{\nu _3 } }}{{P_{\nu _3 } }})(p_{0,\nu _2 }^L p_{1,\nu _1 }^U  - p_{0,\nu _1 }^U p_{1,\nu _2 }^L )\bigg]p_{3,\mu }^L
  \nonumber \\
&&-\bigg[(p_{0,\nu _2 }^L p_{3,\nu _1 }^U  - p_{0,\nu _1 }^U p_{3,\nu _2 }^L )(p_{0,\nu _3 }^U p_{1,\nu _2 }^L  - p_{0,\nu _2 }^L p_{1,\nu _3 }^U ) - (p_{0,\nu _3 }^U p_{3,\nu _2 }^L  - p_{0,\nu _2 }^L p_{3,\nu _3 }^U )(p_{0,\nu _2 }^L p_{1,\nu _1 }^U  - p_{0,\nu _1 }^U p_{1,\nu _2 }^L )\bigg]
  \nonumber \\
&&(\frac{{N_\mu  }}{{P_\mu  }} - p_{0,\mu }^L D_0^L  - p_{1,\mu }^L D_1^L ) \bigg\} \bigg/ \bigg\{ \bigg[(p_{0,\nu _2 }^L p_{2,\nu _1 }^U  - p_{0,\nu _1 }^U p_{2,\nu _2 }^L )(p_{0,\nu _3 }^U p_{1,\nu _2 }^L  - p_{0,\nu _2 }^L p_{1,\nu _3 }^U )
   \nonumber \\
&&- (p_{0,\nu _3 }^U p_{2,\nu _2 }^L  - p_{0,\nu _2 }^L p_{2,\nu _3 }^U )(p_{0,\nu _2 }^L p_{1,\nu _1 }^U  - p_{0,\nu _1 }^U p_{1,\nu _2 }^L )\bigg] p_{3,\mu }^L- \bigg[(p_{0,\nu _2 }^L p_{3,\nu _1 }^U  - p_{0,\nu _1 }^U p_{3,\nu _2 }^L )(p_{0,\nu _3 }^U p_{1,\nu _2 }^L  - p_{0,\nu _2 }^L p_{1,\nu _3 }^U )
   \nonumber \\
&&- (p_{0,\nu _3 }^U p_{3,\nu _2 }^L  - p_{0,\nu _2 }^L p_{3,\nu _3 }^U )(p_{0,\nu _2 }^L p_{1,\nu _1 }^U  - p_{0,\nu _1 }^U p_{1,\nu _2 }^L )\bigg] p_{2,\mu }^L \bigg\}.
\end{eqnarray}
\end{small}

\bibliographystyle{spphys}       
\bibliography{RRDPSwithSourceErrors2}

\end{document}